\documentclass[12pt,a4paper]{article}

\usepackage{authblk}
\usepackage{amsmath,amssymb}
\usepackage{graphicx,psfrag,color}
\usepackage{amsfonts}
\usepackage{epsfig}
\usepackage{times,pstricks,pst-node}
\usepackage{pstricks,pst-node,pst-text,pst-3d}
\usepackage{caption}
\usepackage{setspace}
\usepackage[section]{placeins}

\renewcommand\footnotemark{}

\newcommand{\im}{\mathrm i}

\newcommand{\tr}{\operatorname{Tr}}
\newcommand{\str}{\operatorname{str}}

\newcommand{\ket}[1]{\left|#1\right\rangle}      
\newcommand{\eq}{\begin{equation}}
\newcommand{\en}{\end{equation}}
\newcommand{\bear}{\begin{eqnarray}}
\newcommand{\ear}{\end{eqnarray}}

\begin{document}
\title{Finite size and finite temperature studies of the $osp(1|2)$ spin chain}
\author{T.S. Tavares$^*$\footnote{$^*$ tavares@df.ufscar.br}  and G.A.P. Ribeiro$^{\dagger}$ \footnote{$^{\dagger}$ pavan@df.ufscar.br}}

\affil{Departamento de F\'{i}sica, Universidade Federal de S\~ao Carlos \\ 13565-905 S\~ao Carlos-SP, Brazil}

\date{}
\maketitle
\begin{center}
{\it Dedicated to the memory of Petr Petrovich Kulish}
\end{center}

\setcounter{page}{1}
\thispagestyle{empty}

\begin{abstract}
We study a quantum spin chain invariant by the superalgebra $osp(1|2)$. We derived non-linear integral equations for the row-to-row transfer matrix eigenvalue in order to analyze its finite size scaling behaviour and we determined its central charge. We have also studied the thermodynamical properties of the obtained spin chain via the non-linear integral equations for the quantum transfer matrix eigenvalue. We numerically solved these NLIE and evaluated the specific heat and magnetic susceptibility. The analytical low temperature analysis was performed providing a different value for the effective central charge. The computed values are in agreement  with the numerical predictions in the literature.
\end{abstract}

\newpage

\section{Introduction}

The notion of superalgebras\cite{KAC} attracted a lot of attention and it was soon considered in the context of Yang-Baxter integrability. This resulted in many new solutions of the Yang-Baxter equation\cite{KULISH82,KULISH,BAZHANOV,SALEUR1990,GOULD}.

Nevertheless, the understanding of critical behaviour of integrable spin chains with bosonic and fermionic degrees of freedom was proven to have its own subtleties\cite{MARTINS1994,MARTINS1997,MARTINS2012}. On the one hand, the spin chains invariant by $osp(n|2m)$ superalgebras are conformally invariant\cite{MARTINS1998}, on the other hand the spin chains invariant by $sl(n|m)$ superalgebras seems to be not even relativistic\cite{SALEUR2000}. This unusual behavior includes excitations with zero conformal weights and ground state degeneracy being dominated by logarithmic finite-size corrections\cite{MARTINS1998,SALEUR2005,SALEUR2008}.

In order to better investigate e.g such logarithmic corrections, it could be useful to have additional analytical tools and efficient numerical approach at hand for the case of spin chains invariant by superalgebras. Along these lines, there exist a couple of approaches which avoid the need of solving numerically Bethe ansatz equations for very long lattices. These approaches result in sets of non-linear integral equations. In the case of finite system size, one can obtain under certain analyticity conditions a set of non-linear integral equations for the largest eigenvalue of the transfer matrix as a function of the chain length $L$ (zero temperature)\cite{KLUMPER91a,KLUMPER91b}. Alternatively, one can use the quantum transfer matrix approach to formulate different non-linear integral equations for the finite temperature case (infinity length)\cite{KLUMPER92,DEVEGA0,KLUMPER93}. Both cases allow for accurate numerical results as well as for analytical solution for certain regimes. However, the mentioned approach has the shortcoming that there is no general method to derive such non-linear integral equations. Therefore, this is done on a case by case basis\cite{KLUMPER93,KLUMPER-TJ,HUBBARD,JSUZUKI,DAMERAU,RIBEIRO,TAVARES}.

In this paper we address the case of a spin chain invariant by the simplest orthosympletic superalgebra, the $osp(1|2)$ case. Although this case was considered before by means of the thermodynamical Bethe ansatz\cite{TSUBOI}, the resulting system has an infinite number of equations, which poses problem in the numerical evaluation of physical quantities at very low temperatures. Therefore, we use a different approach which results in a finite number of non-linear integral equations. The equations derived here allow for accurate numerical evaluation of the transfer matrix eigenvalue as a function of system size as well as the thermodynamical potential and related quantities as function of temperature. We have also obtained the central charge analytically.

This paper is organized as follows. First we present the quantum spin chain invariant by $osp(1|2)$ superalgebra. In the section \ref{finitesize} we derive a system of non-linear integral equations for the finite-size corrections to the transfer matrix eigenvalue. The section \ref{qtm} is devoted to the derivation of non-linear integral equations for finite temperature via quantum transfer matrix approach. Our conclusions are given in section \ref{conclusion}.

\section{The $osp(1|2)$ spin chain}\label{osp12}

The theory of quantum integrable models in one-dimension is based on the Yang-Baxter equation\cite{BAXTER-book,KOREPIN-book}. In the context of models with bosonic and fermionic degrees of freedom, the Yang-Baxter equation can be appropriately generalized to its graded version, which with some grading choice accommodates the existence of fermionic degrees\cite{KULISH82,KULISH}. The graded Yang-Baxter equation naturally reads,
\eq
{\cal L}_{12}(\lambda- \mu){\cal L}_{13}(\lambda){\cal L}_{23}(\mu)={\cal L}_{23}(\mu){\cal L}_{13}(\lambda){\cal L}_{12}(\lambda- \mu),
\label{GYBE}
\en
which looks similar to the usual Yang-Baxter equation, as long as we assume that ${\cal L}$ operators act on the super-tensor product of vectors spaces, such that
\eq
\mathcal{L}_{jk}(\lambda)=\sum_{\alpha,\beta,\gamma,\delta} \check{\mathcal{L}}_{\alpha,\gamma}^{\beta ,\delta}(\lambda) e_{\alpha\beta}^{(j)} e_{\gamma\delta}^{(k)}, \qquad e_{\alpha\beta}^{(j)} =\mbox{Id} \stackrel{s}{\otimes} \ldots \stackrel{s}{\otimes} \underbrace{e_{\alpha\beta}}_j\stackrel{s}{\otimes}\ldots \stackrel{s}{\otimes} \mbox{Id},
\en
where $\stackrel{s}{\otimes}$ denotes the super-tensor product\cite{KULISH} and $e_{\alpha\beta}$ is the Weyl basis.

Here we deal with the solution of the Yang-Baxter equation invariant by the $osp(1|2)$ superalgebra\cite{KULISH}, which can be written as
\eq
\mathcal{L}(\lambda)=\lambda \left(\frac{3}{2}-\lambda\right)I+\left(\frac{3}{2}-\lambda\right)P^g+ \lambda E, \label{BWeight}
\en
where $I$ is the identity matrix, $P^g$ is the graded permutation and $E$ is the Temperley-Lieb operators satisfying the Braid-monoid algebra\cite{WADATI}. In the grading $\{p(1),p(2),p(3)\}=\{1,0,1\}$, we have that
\eq
P^g=\sum_{i,j=1}^3 {(-1)}^{p(i) p(j)} e_{i j} \otimes e_{j i}, \qquad E = \sum_{i,j=1}^3 {(-1)}^{\delta_{i 3}+\delta_{j 1}} e_{i j} \otimes e_{4-i~4-j}.
\en

The graded Yang-Baxter equation provides the commutativity property $[T(\lambda),T(\mu)]=0$ of the transfer matrix
\eq
T(\lambda)=\str_{\cal A} \left[{\cal G}_{\cal A} \prod_{j=1}^{\stackrel{\curvearrowleft}{L}}\mathcal{L}_{{\cal A} j}(\lambda)\right],
\label{r-t-r}
\en
where $\str_{\cal A} $ denotes the super-trace over the auxiliary space. For later convenience, we introduced twisted boundary conditions along the horizontal $\left({\cal G}\right)_{j j}={\rm e}^{-\im \eta (j-2)}$. The $\cal L$-operator (\ref{BWeight}) satisfies the following properties:
\begin{align}
  \mbox{Regularity: } & {\cal L}_{12}(0)=a(0) P_{12}^g, \label{regul} \\
  \mbox{Unitarity: } & {\cal L}_{12}(\lambda) {\cal L}_{12}(-\lambda)= a(\lambda) a(-\lambda)I, \label{uni} \\
  \mbox{Time reversal: } & {\cal L}_{12}^{st_1}(\lambda)={\cal L}_{12}^{st_2}(\lambda), \label{time-rev}
\end{align}
where $st_i$ denotes the super-transpose in $i$-th space and $a(\lambda)=(1-\lambda) (\frac{3}{2}-\lambda)$.

Thanks to these properties, we have that the logarithmic derivative of the row-to-row transfer matrix results in a quantum spin chain Hamiltonian, which can be written as,
\bear
{\cal H}&=&-J\frac{d}{d \lambda}\ln{\left(\frac{T(\lambda)}{a^L(\lambda)}\right)}\Bigg|_{\lambda=0} \nonumber\\
&=&J\sum_{j=1}^L\Bigg[ -\sum_{\sigma}\Big( c^\dag_{j+1 \sigma} c_{j \sigma}+c^\dag_{j \sigma} c_{j+1 \sigma}-\frac{2}{3} \mbox{sgn}(\sigma) (c_{j \sigma} c_{j+1 \sigma}+c_{j \sigma}^{\dag} c_{j+1 \sigma}^{\dag}) \nonumber\\
&-&\frac{5}{3}(n_{j \sigma}+n_{j+1 \sigma})\Big)+\frac{1}{3} \vec{S}_j \cdot \vec{S}_{j+1}-\frac{5}{6}\sum_{\sigma \sigma' }( n_{j \sigma} n_{j+1 \sigma'})-\frac{8}{3}\Bigg],
\label{Spinchain}
\ear
where $J=1$, $n_{j \sigma}=c_{j \sigma}^{\dag} c_{j \sigma}$, $S_j^k=\sum_{\sigma \sigma'} S^k_{\sigma \sigma'} c_{j \sigma}^{\dag} c_{j \sigma'}$ ($k=x,~y,~z$) and
$c_{j \sigma}$ are the ``projected'' fermionic operators acting on subspace $\ket{\uparrow}, ~\ket{0}, ~\ket{\downarrow}$ with grading $\{1, ~0, ~1\}$. These operators satisfy exactly the same anti-commutation rules as in the t-J model\cite{CHAO,TAVARES2016}, which prevents double occupation of a single site. The critical properties of this model was firstly studied via finite-size scaling analysis\cite{MARTINS1998}.

On the other hand, we can also study the thermodynamics of the above spin chain via quantum transfer matrix(QTM) approach\cite{MSUZUKI,KLUMPER92,DEVEGA0,KLUMPER93}. This is usually done by mapping the problem of the evaluation of the partition function of the quantum chain $Z=\tr{\left[e^{-\beta {\cal H}}\right]}$ into the evaluation of the partition function of a suitable bidimensional classical vertex model via the Trotter-Suzuki decomposition\cite{MSUZUKI}. The  important object obtained from this decomposition is the so called the quantum transfer matrix,
\eq
t^{QTM}(x)= \frac{1}{{(a(\tau+\im x) a(\tau-\im x))}^{\frac{N}{2}}}\tr_{Q}  \prod_{i=1}^{\frac{N}{2}} {\cal L}_{ 2i-1,Q}(\tau, -\im x) {\cal L}_{2 i,Q}^{st_{\cal A}}(-\im x,-\tau).
\label{qtm-gen}
\en
This matrix possess a number of convenient properties which allows us to determine the grand partition function of model (\ref{Spinchain}) at fixed chemical potential out of its largest eigenvalue. Here $\tau= \frac{\beta}{N}$ introduces the temperature dependence.  We notice that model (\ref{Spinchain}), although possessing real eigenvalues, has a non-hermitian term which also does not conserve particle number. Such term precludes the evaluation of the grand partition function for arbitrary chemical potential, since an extra contribution, proportional to particle number, cannot be introduced as a twist factor $\mathcal{G}$ in QTM, without spoiling integrability.

In the coming sections we are going to derive NLIE for the largest eigenvalue of row-to-row/quantum transfer matrix for arbitrary finite size/temperature. This will allow us to extract the information about the critical behaviour of the quantum spin chain as well as to provide accurate results for largest eigenvalue as a function of the system size and the thermodynamical properties as a function of temperature.

\section{Row-to-Row Transfer Matrix}\label{finitesize}

In this section, we derive the non-linear integral equations that describe the largest eigenvalue of the row-to-row transfer matrix at finite length $L$. Next we investigate the leading finite size correction and determine the central charge.

\subsection{NLIE for the largest eigenvalue at finite system size}

The eigenvalues of the row-to-row transfer matrix (\ref{r-t-r}) was firstly obtained via analytic Bethe ansatz by Kulish \cite{KULISH} and later on by means of algebraic Bethe ansatz\cite{MARTINS1997}. The eigenvalues can be written in the form
\begin{multline}
\Lambda(\lambda)=-{\rm e}^{\im \eta}{(\lambda-1)}^L {\left(\frac{3}{2}-\lambda\right)}^L \prod_{j=1}^n\frac{\lambda-\mu_j+1}{\lambda-\mu_j} \\
+ {\left(\frac{3}{2}-\lambda\right)}^L \lambda^L \prod_{j=1}^n\frac{(\lambda-\mu_j+\frac{1}{2}) (\lambda-\mu_j-1)}{(\lambda-\mu_j-\frac{1}{2})(\lambda-\mu_j)}\\-{\rm e}^{-\im \eta} \lambda^L {\left(\frac{1}{2}-\lambda\right)}^L \prod_{j=1}^n\frac{\lambda-\mu_j-\frac{3}{2}}{\lambda-\mu_j-\frac{1}{2}}=\lambda_1(\lambda)+\lambda_2(\lambda)+\lambda_3(\lambda),
\label{r-t-reig}
\end{multline}
where the Bethe ansatz roots $\mu_j$ must satisfy the Bethe equations
\eq
{\left(\frac{\mu_k-1}{\mu_k}\right)}^L={\rm e}^{-\im \eta} \prod_{\stackrel{j=1}{j \neq k}}^n \frac{(\mu_k-\mu_j+\frac{1}{2})(\mu_k-\mu_j-1)}{(\mu_k-\mu_j+1)(\mu_k-\mu_j-\frac{1}{2})}. \label{BAE}
\en
It is worth to note that although we have three pieces adding up to form the eigenvalue, there is only one type of Bethe ansatz roots. This means that the one level Bethe equations are sufficient to induce pole cancellations in both sums $\lambda_1(\mu)+\lambda_2(\mu)$ and $\lambda_2(\mu)+\lambda_3(\mu)$, resulting in an analytical eigenvalue. Such feature does take place when there is a constraint among the three functions
\eq
\lambda_1(\mu-\frac{1}{4})\lambda_3(\mu+\frac{1}{4})=\lambda_2(\mu-\frac{1}{4}) \lambda_2(\mu+\frac{1}{4}).
\en
This is a consequence of the existence of fusion hierarchy\cite{TSUBOI-fusion} for the transfer matrix (\ref{r-t-r}).

To our further development, we introduce functions $\tilde{\Lambda}(x)= \Lambda(\im x+\frac{3}{4})$, $\phi(x)=x^L$ and $Q(x)=\prod_{j=1}^m (x-x_j)$, where we shift Bethe ansatz roots as $\mu_j=\frac{1}{2}-\im x_j$. The new Bethe ansatz roots $x_j$ are real for the largest eigenvalue, which lies in sector $n=L$. Strictly at $\eta=0$, we are not able to solve Bethe ansatz equations in the largest eigenvalue sector. However, if we slowly change $\eta$ from some finite value to zero, we find that one Bethe ansatz root is moving to infinity along the real axis, which explains the three-fold degeneracy of the largest eigenvalue at $\eta=0$. In this case, a similar behavior appears for the largest eigenvalue in the sector $n=L+1$, when two roots go to infinity, making it equal to the largest eigenvalue at $n=L-1$.

Defining the function
\eq
S(x)=\underbrace{{\rm e}^{\im \frac{\eta}{2}} \phi(x-\frac{\im}{2})\frac{ Q(x+\im) Q(x-\frac{\im}{2})}{Q(x)}}_{s_1(x)}\underbrace{-{\rm e}^{-\im \frac{\eta}{2}} \phi(x+\frac{\im}{2})\frac{ Q(x-\im) Q(x+\frac{\im}{2})}{Q(x)}}_{s_2(x)}, \label{AuxLamb}
\en
\bear
s_1(x)&=&{\rm e}^{\im \frac{\eta}{2}} \phi(x-\frac{\im}{2})\frac{ Q(x+\im) Q(x-\frac{\im}{2})}{Q(x)}, \\
s_2(x)&=&-{\rm e}^{-\im \frac{\eta}{2}} \phi(x+\frac{\im}{2})\frac{ Q(x-\im) Q(x+\frac{\im}{2})}{Q(x)},
\ear
which is entire as long as Bethe ansatz equations (\ref{BAE}) are satisfied, we find
\eq
\tilde{\Lambda}(x)=-{(-1)}^L \left[\frac{S(x+\frac{\im}{4}) S(x-\frac{\im}{4})+\phi(x+\frac{\im}{4}) \phi(x-\frac{\im}{4}) Q(x+\frac{5 \im}{4}) Q(x-\frac{5 \im}{4})}{Q(x+\frac{3 \im}{4}) Q(x-\frac{3 \im}{4})}\right]. \label{fus}
\en
Moreover, Bethe ansatz equations guarantees that the above $S(x)$ function associated to the largest eigenvalue is analytical and non-zero (ANZ) inside the strip $|\Im z|\leq 1$. For $\tilde{\Lambda}(x)$, despite poles cancellations due to the Bethe ansatz equation, there remains two symmetric real zeros. This naturally implies that the function $\tilde{\Lambda}(x)$ is no longer ANZ inside the strip.

\begin{figure}[htb]
\begin{center}
\includegraphics[width=0.6 \linewidth]{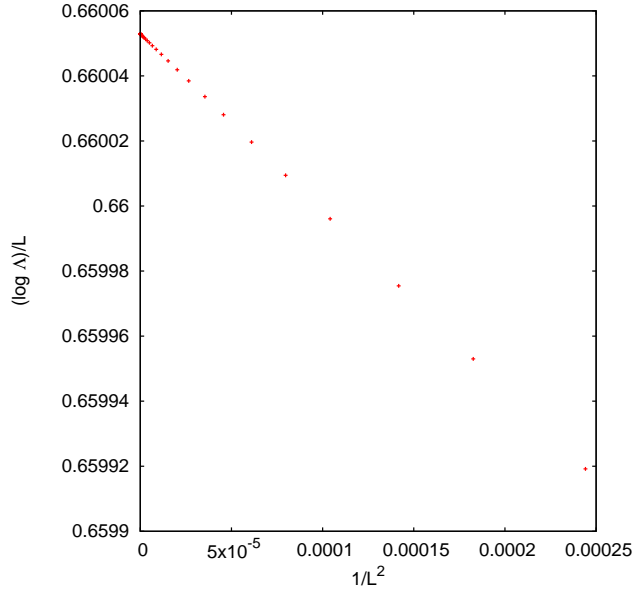}
\end{center}
\caption{Largest eigenvalue $\frac{\log \tilde{\Lambda}(0)}{L}$ vs. $\frac{1}{L^2}$ for $\eta=0.6$, $L=80,~160,\ldots,4000$. While the $y$-intercept tends to the exact value $D_Y(0)+D_{\Lambda}(0)$, the slope provides a numerical approximation to the central charge.}
\label{fig00}
\end{figure}

Instead of directly solving Bethe ansatz equations for very large system sizes, one may try to solve the functional problem that is to find a function of form (\ref{AuxLamb}) that also possess the above analytical non-zero strip.  In the functional problem  we define auxiliary functions in a product form
\begin{align}
\mathfrak{b}(x)&=\frac{s_1(x+\im \alpha)}{s_2(x+\im \alpha)}=-{\rm e}^{\im \eta} \frac{\phi(x-\frac{\im}{2}+\im \alpha) Q(x+\im+\im \alpha) Q(x-\frac{\im}{2}+\im \alpha)}{\phi(x+\frac{\im}{2}+\im \alpha) Q(x-\im+\im \alpha) Q(x+\frac{\im}{2}+\im \alpha)},\nonumber\\
\bar{\mathfrak{b}}(x)&=\frac{s_2(x-\im \alpha)}{s_1(x-\im \alpha)}=-{\rm e}^{-\im \eta} \frac{\phi(x+\frac{\im}{2}-\im \alpha) Q(x-\im-\im \alpha) Q(x+\frac{\im}{2}-\im \alpha)}{\phi(x-\frac{\im}{2}-\im \alpha) Q(x+\im-\im \alpha) Q(x-\frac{\im}{2}-\im \alpha)}, \label{Auxdefi}
\end{align}
with $0<\alpha<\frac{1}{2}$. In addition, these functions have the constant asymptotic limit
\eq
\lim_{x \rightarrow \pm \infty} \mathfrak{b}(x)= -{\rm e}^{\im \eta}~~~~\lim_{x \rightarrow \pm \infty} \bar{\mathfrak{b}}(x)= -{\rm e}^{-\im \eta}.
\label{Asympt}
\en
This allows us to Fourier transform the logarithm derivative of these functions, likewise the closely related functions defined by
\begin{align}
\mathfrak{B}(x)&=1+\mathfrak{b}(x)=\frac{-{\rm e}^{\im \frac{\eta}{2}} S(x+\im \alpha) Q(x+\im \alpha)}{\phi (x+\frac{\im}{2}+\im \alpha)Q(x-\im+\im \alpha) Q(x+\frac{\im}{2}+\im \alpha)}, \nonumber\\
\bar{\mathfrak{B}}(x)&=1+\bar{\mathfrak{b}}(x)=\frac{-{\rm e}^{-\im \frac{\eta}{2}} S(x-\im \alpha) Q(x-\im \alpha)}{\phi (x-\frac{\im}{2}-\im \alpha)Q(x+\im-\im \alpha) Q(x-\frac{\im}{2}-\im \alpha)}. \label{Auxcap}
\end{align}
By solving (\ref{Auxcap}) for $Q(x)$ and $S(x)$ in the Fourier space and replacing the result in Fourier transform of (\ref{Auxdefi}), transforming back to the real space, and integrating the result from $-\infty$ to $x$, we obtain that
\begin{align}
\log \mathfrak{b}(x)&=-F \ast \log \mathfrak{B}(x)+F\ast \log \bar{\mathfrak{B}}(x+2 \alpha \im)+\im (L~{\cal D}(x+\im \alpha)+\eta-\pi \mbox{sgn}(\eta)), \nonumber\\
\log \bar{\mathfrak{b}}(x)&=~~~F \ast \log \mathfrak{B}(x-2 \alpha \im)-F\ast \log \bar{\mathfrak{B}}(x)-\im (L~{\cal D}(x-\im \alpha)+\eta-\pi \mbox{sgn}(\eta)), \label{NLIEFinite}
\end{align}
where we denote the convolution $(f \ast g)(x)=\frac{1}{2 \pi}\int_{-\infty}^{\infty} f(x-s)g(s){\rm d}s$,
\begin{equation}
{\cal D}(x)=-\im \log \left[\frac{{\rm e}^{-\frac{2 \pi x}{3}}+{\rm e}^{\frac{\pi \im}{3}}}{{\rm e}^{-\frac{2 \pi x}{3}}-{\rm e}^{\frac{\pi \im}{3}}}\right]-\im \log \left[\frac{{\rm e}^{-\frac{2 \pi x}{3}}-{\rm e}^{-\frac{\pi \im}{3}}}{{\rm e}^{-\frac{2 \pi x}{3}}+{\rm e}^{-\frac{\pi \im}{3}}}\right], \label{driving}
\end{equation}
and
\begin{align}
F(x)&=\int_{-\infty}^{\infty}\frac{{\rm e}^{\frac{-|k|}{2}}-{\rm e}^{-|k|}}{1+{\rm e}^{-|k|}-{\rm e}^{\frac{-|k|}{2}}}{\rm e}^{\im k x}{\rm d}k \nonumber\\
&=\frac{{\rm d}}{\im {\rm d}x}\log \left[\frac{\Gamma(\frac{1}{6}-\frac{\im x}{3})\Gamma(\frac{1}{2}+\frac{\im x}{3})\Gamma(\frac{2}{3}+\frac{\im x}{3})\Gamma(1-\frac{\im x}{3})}{\Gamma(\frac{1}{6}+\frac{\im x}{3})\Gamma(\frac{1}{2}-\frac{\im x}{3})\Gamma(\frac{2}{3}-\frac{\im x}{3})\Gamma(1+\frac{\im x}{3})}\right]. \label{F(x)}
\end{align}
The eigenvalue $S(x)$ becomes
\eq
\log S(x)=G_1 \ast \log \mathfrak{B}(x-\im \alpha)+\bar{G}_1\ast \log \bar{\mathfrak{B}}(x+\im \alpha)+L D_S(x)+\frac{\pi \im}{2} \mbox{sgn}(\eta),
\en
where
\eq
D_S(x)=\log(x^2+\frac{1}{4})+\log \left[\frac{\Gamma(\frac{1}{6}+\frac{\im x}{3})\Gamma(\frac{1}{6}-\frac{\im x}{3})\Gamma(1+\frac{\im x}{3})\Gamma(1-\frac{\im x}{3})}{\Gamma(\frac{1}{2}+\frac{\im x}{3})\Gamma(\frac{1}{2}-\frac{\im x}{3})\Gamma(\frac{2}{3}+\frac{\im x}{3})\Gamma(\frac{2}{3}-\frac{\im x}{3})}\right]\nonumber,
\en
and
\eq
G_1(x)=\int_{-\infty}^{\infty}{\rm e}^{\im k x} \hat{G}_1(k){\rm d}k\qquad \text{with}~~\hat{G}_1(k)=\begin{cases}
\frac{{\rm e}^{-k}}{1+{\rm e}^{-k}-{\rm e}^{-\frac{k}{2}}}, & \text{if}~k \geq 0\\
\frac{1-{\rm e}^{\frac{k}{2}}}{1+{\rm e}^{k}-{\rm e}^{\frac{k}{2}}}, & \text{if}~k <0
\end{cases}\nonumber.
\en
Now we want to use the hierarchy relation in order to obtain $\tilde{\Lambda}(x)$. If we introduce the $Y$-system as
\begin{align}
y(x)&=\frac{{(-1)}^L \tilde{\Lambda}(x) Q(x-\frac{3 \im}{4})Q(x+\frac{3 \im}{4}) }{\phi(x-\frac{\im}{4}) \phi(x+\frac{\im}{4}) Q(x+\frac{5 \im}{4})Q(x-\frac{5 \im}{4})},\nonumber\\
Y(x)&=\frac{-S(x+\frac{\im}{4}) S(x-\frac{\im}{4}) }{\phi(x-\frac{\im}{4}) \phi(x+\frac{\im}{4}) Q(x+\frac{5 \im}{4})Q(x-\frac{5 \im}{4})},
\end{align}
then the fusion hierarchy (\ref{fus}) is simply written as $Y(x)=1+y(x)$. Therefore the previous calculation also allow us to determine $Y(x)$ function. Fourier transforming the logarithm derivative of $Y(x)$ and proceeding as before, we find
\eq
\log Y(x)=G_2 \ast \log \mathfrak{B}(x-\im \alpha)+\bar{G}_2 \ast \log \bar{\mathfrak{B}}(x+\im \alpha) +L D_Y(x), \label{Y}
\en
where $D_Y(x)=-\im ({\cal D}(x+\frac{\im}{4})-{\cal D}(x-\frac{\im}{4}))$, and
\eq
G_2(x)=\int_{-\infty}^{\infty}{\rm e}^{\im k x} \hat{G}_2(k){\rm d}k\qquad \text{with}~~\hat{G}_2(k)=\begin{cases}
\frac{{\rm e}^{-\frac{3 k}{4}}}{1+{\rm e}^{-k}-{\rm e}^{-\frac{k}{2}}}, & \text{if}~k \geq 0\\
\frac{{\rm e}^{-\frac{k}{4}}(1+{\rm e}^{\frac{3 k}{2}}-{\rm e}^{k})}{1+{\rm e}^{k}-{\rm e}^{\frac{k}{2}}}, & \text{if}~k <0
\end{cases}\nonumber.
\en
We note that in (\ref{Y}) one has to regularize $G_2(x)$ by choosing $\alpha \geq \frac{1}{4}$.
Once the non linear integral equations are solved, we find $Y(x)$ and immediately $y(x)$, because of the obvious constraint. Also, Fourier transforming the logarithm derivative of $y(x)$ permits us to relate this function to the eigenvalue we are interested in. The final result is given as
\eq
\log \tilde{\Lambda}(x)=G_3 \ast \log \mathfrak{B}(x-\im \alpha)-G_3 \ast \log \bar{\mathfrak{B}}(x+\im \alpha)+\log y(x)+ L D_{\Lambda}(x)+ \pi \im \mbox{Mod}(L,2),
\en
where
\begin{multline}
D_{\Lambda}(x)= \log(x+\frac{\im}{4})+\log(x-\frac{\im}{4}) \\-\log\left[\frac{\Gamma(\frac{2}{3}+\frac{1}{12}+\frac{\im x}{3})\Gamma(\frac{2}{3}+\frac{1}{12}-\frac{\im x}{3})\Gamma(\frac{5}{6}+\frac{1}{12}+\frac{\im x}{3})\Gamma(\frac{5}{6}+\frac{1}{12}-\frac{\im x}{3})}{\Gamma(\frac{7}{6}+\frac{1}{12}+\frac{\im x}{3})\Gamma(\frac{7}{6}+\frac{1}{12}-\frac{\im x}{3})\Gamma(\frac{1}{3}+\frac{1}{12}+\frac{\im x}{3})\Gamma(\frac{1}{3}+\frac{1}{12}-\frac{\im x}{3})}\right],
\nonumber
\end{multline}
and
\eq
G_3(x)=\int_{-\infty}^{\infty}{\rm e}^{\im k x} \hat{G}_3(k){\rm d}k\qquad \text{with}~~\hat{G}_3(k)=\begin{cases}
\frac{-{\rm e}^{-\frac{3 k}{4}} (1-{\rm e}^{-\frac{k}{2}})}{1+{\rm e}^{-k}-{\rm e}^{-\frac{k}{2}}}, & \text{if}~k \geq 0\\
\frac{{\rm e}^{\frac{3 k}{4}} (1-{\rm e}^{\frac{k}{2}})}{1+{\rm e}^{k}-{\rm e}^{\frac{k}{2}}}, & \text{if}~k <0
\end{cases}.\nonumber
\en
The numerical solution of NLIE's (\ref{NLIEFinite}) allow us to compute the largest eigenvalue. In Figure \ref{fig00} we plot $\frac{\log \tilde{\Lambda}(0)}{L}$ against $\frac{1}{L^2}$ for $\eta=0.6$. The linear shape highlights conformal behavior, where the slope is proportional to the central charge. The numerical results obtained here are consistent with the analytical expressions to be described in the next section.

\subsection{Leading finite size correction and central charge}

Usually one makes use of a different set of auxiliary functions to derive the NLIE when the eigenvalue of interest is not the most fundamental one in fusion hierarchy, see for example \cite{JSUZUKI}. Our choice in the previous section reduces the number of equations, although having the shortcoming that the eigenvalue is obtained rather indirectly. After solving (\ref{NLIEFinite}), the function $Y(x)$ is readily evaluated. This allows us to evaluate $y(x)$ and finally the largest eigenvalue $\tilde{\Lambda}(x)$.

Now we will consider the finite size analysis of the largest eigenvalue of the transfer matrix, which provides us the evaluation of the central charge \cite{KLUMPER91a,KLUMPER91b}.

We assume that
\eq
\log Y (x)= L f +\sigma(L),
\en
where $\lim_{L \rightarrow \infty} \frac{\sigma(L)}{L} \rightarrow 0$ and $f>0$. This is indeed true since $D_Y(x)>0$ for all $x$. Using the relation $y(x)=Y(x)-1$, we find
\eq
\log y (x)= L f+ \sigma+\log\left(1-{\rm e}^{-L f-\sigma}\right)= L f+\sigma-\underbrace{\sum_{n=1}^{\infty}\frac{{\rm e}^{-n (L f+\sigma)}}{n}}_{\delta}.
\en
Notice that $\lim_{L \rightarrow \infty} L^{\gamma} \delta=0$ for any $\gamma \geq 0$. Therefore $\log y(x)$ and $\log Y(x)$ possess the same algebraic behavior and the difference $\delta=\log Y (x)-\log y(x)$ decays exponentially with system size $L$. Because of this, we find
\begin{multline}
\log \tilde{\Lambda}(x)=L (D_Y(x)+D_{\Lambda}(x))-\delta+ \\
\left(G_2+G_3\right)\ast \log \mathfrak{B}(x-\im \alpha)+\left(\bar{G}_2-G_3\right)\ast \log \bar{\mathfrak{B}}(x+\im \alpha), \label{autosx}
\end{multline}
and therefore all contributions algebraically decaying on system size comes from convolutions. Now we observe that kernels in (\ref{autosx}) is related to the driving-term of (\ref{NLIEFinite})
\begin{align}
G_2(x-\im \alpha)+G_3(x-\im \alpha)&={\cal D}'\left(x+ \im\left(\frac{3}{4}-\alpha\right)\right) \nonumber \\
\bar{G}_2(x+\im \alpha)-G_3(x+\im \alpha)&= {\cal D}'\left(x- \im \left(\frac{3}{4}-\alpha\right)\right).
\end{align}
All non-exponential decaying corrections is calculated from
\eq
\mbox{cor}(x)={\cal D}'\ast \log \mathfrak{B}(x+\frac{\im}{2})+{\cal D}' \ast \log \bar{\mathfrak{B}}(x-\frac{\im}{2}),
\en
where we fixed $\alpha=\frac{1}{4}$.

Let $s$ be integration variable of convolutions. After the change of variables $s=\pm(s'+\frac{3}{2\pi} \log L)$, we find
\begin{multline}
\mbox{cor}(x)=\int_{-\frac{3}{2 \pi} \log L}^{\infty} {\cal D}_+'(x-s'+\frac{\im}{2})  l\mathfrak{B}^+(s')
+ {\cal D}_-'(x+s'+\frac{\im}{2})  l\mathfrak{B}^-(s')
{\rm d}s'+\\
\int_{-\frac{3}{2 \pi} \log L}^{\infty} {\cal D}_+'(x-s'-\frac{\im}{2})  l\bar{\mathfrak{B}}^+(s')
+ {\cal D}_-'(x+s'-\frac{\im}{2})  l\bar{\mathfrak{B}}^-(s')
{\rm d}s',
\end{multline}
where we have defined
\begin{align}
l \mathfrak{b}^{\pm}(x)&:=\log \mathfrak{b}(\pm(x+\frac{3}{2 \pi}\log L))~~~~l \mathfrak{B}^{\pm}(x):=\log \mathfrak{B}(\pm(x+\frac{3}{2 \pi}\log L)), \nonumber\\
l \bar{\mathfrak{b}}^{\pm}(x)&:=\log \bar{\mathfrak{b}}(\pm(x+\frac{3}{2 \pi}\log L))~~~~l \bar{\mathfrak{B}}^{\pm}(x):=\log \bar{\mathfrak{B}}(\pm(x+\frac{3}{2 \pi}\log L)),\nonumber\\
{\cal D}_{\pm}'(x)&:={\cal D}'(x\mp\frac{3}{2 \pi} \log L)\approx \pm \frac{4}{L}{\rm e}^{\pm\frac{2 \pi x}{3}} \sin\frac{\pi}{3}~~~~(\mbox{Mod}~2 \pi ).\label{Aprox}
\end{align}
Because of approximation (\ref{Aprox}), we have
\begin{multline}
\mbox{cor}(x)=\frac{4}{3 L} \sin(\frac{\pi}{3}) \Bigg\{ {\rm e}^{\frac{2 \pi x}{3}} \int_{-\frac{3}{2 \pi} \log L}^{\infty} {\rm e}^{-\frac{2 \pi s}{3}} \left({\rm e}^{\frac{\pi \im}{3}} l \mathfrak{B}^{+}(s)+{\rm e}^{\frac{-\pi \im}{3}} l \bar{\mathfrak{B}}^{+}(s) \right){\rm d}s+\\
{\rm e}^{\frac{-2 \pi x}{3}} \int_{-\frac{3}{2 \pi} \log L}^{\infty} {\rm e}^{-\frac{2 \pi s}{3}} \left({\rm e}^{\frac{-\pi \im}{3}} l \mathfrak{B}^{-}(s)+{\rm e}^{\frac{\pi \im}{3}} l \bar{\mathfrak{B}}^{-}(s) \right){\rm d}s
\Bigg\}. \label{Correction}
\end{multline}

Now we use $x=s\pm \frac{3}{2 \pi}\log L$ in the (\ref{NLIEFinite})
\begin{align}
\l\mathfrak{b}^{\pm}(s)&=-F \ast l \mathfrak{B}^{\pm}(s)+F\ast l\bar{\mathfrak{B}}^{\pm}(s+\frac{\im}{2})-4 \sin(\frac{\pi}{3}){\rm e}^{-\frac{2 \pi s}{3}\pm \frac{\pi \im}{3}}+\im (\eta-\pi \mbox{sgn}(\eta)), \nonumber\\
\l\bar{\mathfrak{b}}^{\pm}(s)&=F \ast l\mathfrak{B}^{\pm}(s-\frac{\im}{2})-F\ast \log \bar{\mathfrak{B}}^{\pm}(s)-4 \sin(\frac{\pi}{3}){\rm e}^{-\frac{2 \pi s}{3}\mp \frac{\pi \im}{3}}-\im (\eta-\pi \mbox{sgn}(\eta)),
\end{align}
and construct the following quantities
\eq
\Delta_{\pm}:=\int_{-\frac{3}{2 \pi} \log L \approx -\infty}^{\infty}
~{\begin{bmatrix}l \mathfrak{B}^{\pm}(s)\\
               l \bar{\mathfrak{B}}^{\pm}(s)
\end{bmatrix}}^{t}\cdot
\begin{bmatrix}l \mathfrak{b}^{\pm\prime}(s)\\
               l \bar{\mathfrak{b}}^{\pm\prime}(s)
\end{bmatrix}
-{\begin{bmatrix}l \mathfrak{B}^{\pm\prime}(s)\\
               l \bar{\mathfrak{B}}^{\pm\prime}(s)
\end{bmatrix}}^{t}\cdot
\begin{bmatrix}l \mathfrak{b}^{\pm}(s)\\
               l \bar{\mathfrak{b}}^{\pm}(s)
\end{bmatrix}{\rm d}s\label{Deltapm0}.
\en
Because of the symmetry property of the kernel $F(s)=F(-s)$, the convolution terms are exactly canceled in $\Delta_{\pm}$. We find
\begin{multline}
\Delta_{\pm}\approx\frac{16 \pi}{3} \sin\left(\frac{\pi}{3}\right) \int_{-\frac{3}{2 \pi} \log L}^{\infty} {\rm e}^{-\frac{2 \pi s}{3}}\left({\rm e}^{\pm \frac{\pi \im}{3}}l \mathfrak{B}^{\pm}(s)+{\rm e}^{\mp\frac{\pi \im}{3}}l \bar{\mathfrak{B}}^{\pm}(s)\right){\rm d}s+\\ 4 \sin\left(\frac{\pi}{3}\right) {\rm e}^{-\frac{2 \pi s}{3}}\left[{\rm e}^{\frac{\pm \pi \im}{3}}l \mathfrak{B}^{\pm}(s) +{\rm e}^{\frac{\mp \pi \im}{3}}l \bar{\mathfrak{B}}^{\pm}(s)-\im(\eta-\pi \mbox{sgn}(\eta)) ( l \mathfrak{B}^{\pm}(s)- l \bar{\mathfrak{B}}^{\pm}(s))\right]_{-\frac{3}{2 \pi}\log L}^{\infty}.
\label{Deltapm}
\end{multline}

Now we remember that $l \mathfrak{B}^{\pm}(-\frac{3}{2 \pi} \log L)=\log \mathfrak{B}(0)$ and similarly for $l \bar{\mathfrak{B}}^{\pm}$. From definition of auxiliary functions (\ref{Auxdefi}) and Bethe root pattern (along the real axis), one can estimate
\eq
|\mathfrak{b}(0)|,~|\bar{\mathfrak{b}}(0)| \leq {\left(\frac{1}{2 \sqrt{2}}\right)}^{L}, \label{Zeropoint}
\en
which, together with asymptotic limit (\ref{Asympt}), allow us to compute the second line of (\ref{Deltapm}) for large system size. Performing the change of variables $z=\mathfrak{b}(x), \bar{\mathfrak{b}}(x)$ in (\ref{Deltapm0}), we find
\eq
\Delta_{\pm}= 2 \left(L_+(\mathfrak{b}(\pm \infty))+L_+(\bar{\mathfrak{b}}(\pm \infty))\right), \label{Deltapm1}
\en
where $L_+(v)=\frac{1}{2}\int_0^{v}\frac{\log(1+z)}{z}-\frac{\log(z)}{1+z}{\rm d}z$ is the Rogers dilogarithm function satisfying the functional relation\cite{LEWIN}
\eq
L_+(v)+L_+(1/v)=\frac{\pi^2}{6}. \label{Functional}
\en
Therefore putting together (\ref{Asympt}), (\ref{Correction}), (\ref{Deltapm}), (\ref{Zeropoint}), (\ref{Deltapm1}) and (\ref{Functional}), we find
\eq
\lim_{L \rightarrow \infty}\frac{6 L \mbox{cor}(x)}{\pi}=\cosh\left( \frac{2 \pi x}{3}\right) \left(1-3 {\left(1-\frac{|\eta|}{\pi}\right)}^2\right),
\en
 hence the central charge is $c=1-3 {\left(1-\frac{|\eta|}{\pi}\right)}^2$, providing $c=-2$ in the limit $\eta \rightarrow 0$\cite{MARTINS1998}. This result is in agreement with the numerical solution of the non-linear integral equations shown in Figure \ref{fig00}.

\section{Quantum Transfer Matrix}\label{qtm}
\subsection{NLIE for the largest eigenvalue at finite temperature}

The row-to-row transfer matrix studied before, is fully invariant by superalgebra, since super-tensor products and super-traces were taken. On the other hand, in the course of the evaluation of the partition function of the quantum chain $Z=\tr{\left[e^{-\beta {\cal H}}\right]}$, the partition function itself is mapped into a bidimensional classical vertex model on the torus via the Trotter-Suzuki decomposition\cite{MSUZUKI}.
In this case, the quantum transfer matrix is the central object, however, due to this mapping we see that the QTM is written in terms of normal trace along the vertical (quantum) direction due to the definition of the partition function. Therefore, this implies that, besides the naturally different vacuum expectations which appears in the eigenvalues expression, one has no signs due to the super-trace in comparison with the row-to-row case. The final expression for the eigenvalues of the quantum transfer matrix reads,
\begin{multline}
\Lambda^{QTM}(x)= {\left[\frac{(x+\im \tau) (x+\im \tau-\frac{\im}{2})}{(x+\im \tau-\im)(x+\im \tau-\frac{3 \im}{2})}\right]}^{\frac{N}{2}} \prod_{j=1}^n \frac{x-x_j-\im}{x-x_j}\\+{\left[\frac{(x+\im \tau) (x-\im \tau)}{(x+\im \tau-\im)(x-\im \tau+\im)}\right]}^{\frac{N}{2}} \prod_{j=1}^n \frac{(x-x_j-\frac{\im}{2}) (x-x_j+\im)}{(x-x_j) (x-x_j+\frac{\im}{2})}\\ + {\left[\frac{(x-\im \tau) (x-\im \tau+\frac{\im}{2})}{(x-\im \tau+\im)(x-\im \tau+\frac{3 \im}{2})}\right]}^{\frac{N}{2}} \prod_{j=1}^n \frac{x-x_j+\frac{3\im}{2}}{x-x_j+\frac{\im}{2}}=\lambda_1(x)+\lambda_2(x)+\lambda_3(x),
\end{multline}
where the Bethe ansatz roots $x_k$ satisfy the system of non-linear equations
\eq
{\left[\frac{(x_k+\im \tau-\frac{\im}{2}) (x_k-\im \tau+\im)}{(x_k+\im \tau-\frac{3\im}{2}) (x_k-\im \tau)}\right]}^{\frac{N}{2}}= -\prod_{\stackrel{j=1}{j \neq k}}^n \frac{(x_k-x_j-\frac{\im}{2}) (x_k-x_j+\im)}{(x_k-x_j+\frac{\im}{2}) (x_k-x_j-\im)}
\label{BAEQTM}
\en
Once again, Bethe equations imply simultaneous poles cancellations of $\lambda_1(x)+\lambda_2(x)$ and $\lambda_2(x)+\lambda_3(x)$. This is a consequence of
\eq
\lambda_1(x+\frac{\im}{4})\lambda_3(x-\frac{\im}{4})=\lambda_2(x+\frac{\im}{4}) \lambda_2(x-\frac{\im}{4}),
\en
due to the fusion hierarchy\cite{TSUBOI}. Therefore, we may define the eigenvalue
\eq
S^{QTM}(x)=\frac{\Phi_+(x-\frac{\im}{4}) Q(x-\im) Q(x+\frac{\im}{2})}{\Phi_+(x-\frac{5 \im}{4})Q(x)} +
\frac{\Phi_-(x+\frac{\im}{4}) Q(x+\im) Q(x-\frac{\im}{2})}{\Phi_-(x+\frac{5 \im}{4})Q(x)},
\en
whose poles in Bethe ansatz roots $x_k$ become removable singularities because of Bethe equations. Hence, fusion hierarchy may be implemented through $Y$-system  where
\begin{align}
y(x)&=\frac{\Phi_+(x-\frac{3 \im}{2}) \Phi_-(x+\frac{3 \im}{2}) Q(x-\frac{3 \im}{4})Q(x+\frac{3 \im}{4}) \Lambda^{QTM}(x)}{\Phi_+(x-\frac{\im}{2}) \Phi_-(x+\frac{\im}{2}) Q(x+\frac{5 \im}{4})Q(x-\frac{5 \im}{4})},\nonumber\\
Y(x)&=\frac{\Phi_+(x-\frac{3 \im}{2}) \Phi_-(x+\frac{3 \im}{2}) S^{QTM}(x+\frac{\im}{4}) S^{QTM}(x-\frac{\im}{4})}{\Phi_+(x-\frac{\im}{2}) \Phi_-(x+\frac{\im}{2}) Q(x+\frac{5 \im}{4})Q(x-\frac{5 \im}{4})},
\end{align}
and $Y(x)=1+y(x)$. Now we are ready to study analytical hypotheses to transform Bethe ansatz equations into non-linear integral equations.

Despite the resemblances, the analyticity hypotheses for the largest QTM eigenvalue are very different from the row-to-row case. The Bethe ansatz roots related to the largest eigenvalue appear in complex conjugate pairs $x_k^c \pm \im \alpha_k$, where $x_k^c$ is the real center and $\alpha_k$ is the imaginary part. It is remarkable that $\alpha_k$ has a strong dependence on temperature, being very close to $\frac{1}{4}$ for small $\beta$ and very close to zero for $\beta \gg 1$. In the Trotter limit, however, $\alpha_k \simeq \frac{1}{4}$ for any finite $\beta$. Besides, eigenvalues $S^{QTM}(x)$ and $\Lambda^{QTM}(x)$ have an analytical non zero strip at least containing $|\Im z|\leq \frac{1}{2}$.

Because of these differences, it is not possible to use the same auxiliary functions of the row-to-row case as built from $S^{QTM}(x)$ blocks. Instead, we use similar auxiliary functions of \cite{JSUZUKI,RIBEIRO}
\begin{align}
\mathfrak{b}(x)&=\frac{\lambda_3(x+\frac{\im}{4})}{\lambda_1(x+\frac{\im}{4})+\lambda_2(x+\frac{\im}{4})}=\frac{\Phi_-(x+\frac{3 \im}{4}) \Phi_-(x+\frac{\im}{4}) \Phi_+(x-\frac{3 \im}{4}) Q(x+\frac{3 \im}{2})}{\Phi_-(x+\frac{7 \im}{4}) \Phi_-(x+\frac{5 \im}{4}) \Phi_+(x+\frac{\im}{4}) S^{QTM}(x)},\nonumber\\
\bar{\mathfrak{b}}(x)&=\frac{\lambda_1(x-\frac{\im}{4})}{\lambda_2(x-\frac{\im}{4})+\lambda_3(x-\frac{\im}{4})}=\frac{\Phi_+(x-\frac{3 \im}{4}) \Phi_+(x-\frac{\im}{4}) \Phi_-(x+\frac{3 \im}{4}) Q(x-\frac{3 \im}{2})}{\Phi_+(x-\frac{7 \im}{4}) \Phi_+(x-\frac{5 \im}{4}) \Phi_-(x-\frac{\im}{4}) S^{QTM}(x)},\nonumber\\
y_c(x)&=\frac{1}{y(x)}=\frac{\Phi_+(x-\frac{\im}{2}) \Phi_-(x+\frac{\im}{2}) Q(x+\frac{5 \im}{4})Q(x-\frac{5 \im}{4})}{\Phi_+(x-\frac{3 \im}{2}) \Phi_-(x+\frac{3 \im}{2}) Q(x-\frac{3 \im}{4})Q(x+\frac{3 \im}{4}) \Lambda^{QTM}(x)},
\label{LowercaseQTM}
\end{align}
where we have performed a particle-hole conjugation on all functions, including $y(x)$ whose transformed counterpart is $y_c(x)$. We also introduce simply related functions
 \begin{align}
\mathfrak{B}(x)&=\mathfrak{b}(x) +1 &=& \frac{\Phi_+(x-\frac{3 \im}{4}) Q(x+\frac{\im}{2}) \Lambda^{QTM}(x+\frac{\im}{4})}{\Phi_+(x+\frac{\im}{4}) S^{QTM}(x)},\nonumber\\
\bar{\mathfrak{B}}(x)&=\bar{\mathfrak{b}}(x)+1&=& \frac{\Phi_-(x+\frac{3 \im}{4}) Q(x-\frac{\im}{2}) \Lambda^{QTM}(x-\frac{\im}{4})}{\Phi_-(x-\frac{\im}{4}) S(x)},\nonumber \\
Y_c(x)&=y_c(x)+1&=&\frac{S^{QTM}(x+\frac{\im}{4}) S^{QTM}(x-\frac{\im}{4})}{Q(x-\frac{3 \im}{4})Q(x+\frac{3 \im}{4}) \Lambda^{QTM}(x)}. \label{CapitalQTM}
\end{align}

Solving (\ref{CapitalQTM}) for $Q(x)$, $S^{QTM}(x)$ and $\Lambda^{QTM}(x)$ in Fourier space, replacing the result in (\ref{LowercaseQTM}), transforming back to real space and integrating from $-\infty$ to $x$, we finally obtain that
\eq
\begin{bmatrix}
\log \mathfrak{b}(x)\\
\log \bar{\mathfrak{b}}(x)\\
\log y_c(x)
\end{bmatrix}=
-\begin{bmatrix}
F_1 & F_2 & F_3\\
\bar{F}_2 & F_1 & \bar{F}_3\\
\bar{F}_3 & F_3 & F_4
\end{bmatrix} \ast
\begin{bmatrix}
\log \mathfrak{B}(x)\\
\log \bar{\mathfrak{B}}(x)\\
\log Y_c(x)
\end{bmatrix}
-\beta
\begin{bmatrix}
{\cal D}'(x-\frac{\im}{4})\\
{\cal D}'(x+\frac{\im}{4})\\
{\cal D}'(x)
\end{bmatrix}
\en 
where $F_1(x)=F(x)$ as in equation (\ref{F(x)}), and $F_j(x)=\int_{-\infty}^{\infty}{\rm e}^{\im k x} \hat{F}_j(k){\rm d }k$ with
\begin{align}
F_2(k)&=
\begin{cases}
\frac{{\rm e}^{-\frac{k}{2}}-{\rm e}^{-k}+{\rm e}^{- 2 k}}{1+{\rm e}^{-k}-{\rm e}^{-\frac{k}{2}}}, & \text{if}~k \geq 0\\
\frac{{\rm e}^{\frac{k}{2}}}{1+{\rm e}^{k}-{\rm e}^{\frac{k}{2}}}, & \text{if}~k <0
\end{cases}\nonumber.
\\
F_3(k)&={\rm e}^{-\frac{k}{4}} F_1(k)+{\rm e}^{\frac{k}{4}} F_2(k), \nonumber
\\
F_4(k)&={\rm e}^{\frac{k}{4}} F_3(k)+{\rm e}^{-\frac{k}{4}} F_3(-k)+1, \nonumber
\end{align}
and ${\cal D}'(x)$ is defined as before, see (\ref{driving}).

Finally the thermodynamical potential $f$ is obtained at $x=0$ from the following quantity
\eq
-\frac{1}{\beta} \log \Lambda^{QTM}(x)= e(x)-\frac{1}{\beta} \left({\cal D}'\ast \log \mathfrak{B}(x+\frac{\im}{4})+{\cal D}'\ast \log \bar{\mathfrak{B}}(x-\frac{\im}{4})+{\cal D}'\ast \log Y_c(x)\right),
\en 
such that $f=-\frac{1}{\beta} \log \Lambda^{QTM}(x=0)$ and
\eq
e(x)=-\im \frac{\rm d}{{\rm d}x} \log \left[\frac{\Gamma(\frac{5}{6}-\frac{\im x}{3})\Gamma(1-\frac{\im x}{3})\Gamma(\frac{1}{3}+\frac{\im x}{3})\Gamma(\frac{1}{2}+\frac{\im x}{3})}{\Gamma(\frac{5}{6}+\frac{\im x}{3})\Gamma(1+\frac{\im x}{3})\Gamma(\frac{1}{3}-\frac{\im x}{3})\Gamma(\frac{1}{2}-\frac{\im x}{3})}\right],
\nonumber
\en
which at $x=0$ gives the ground-state energy $e_{gs}=e(0)$ per lattice size.

\begin{figure}[htb]
\begin{minipage}{0.5\linewidth}
\begin{center}
\includegraphics[width=\linewidth]{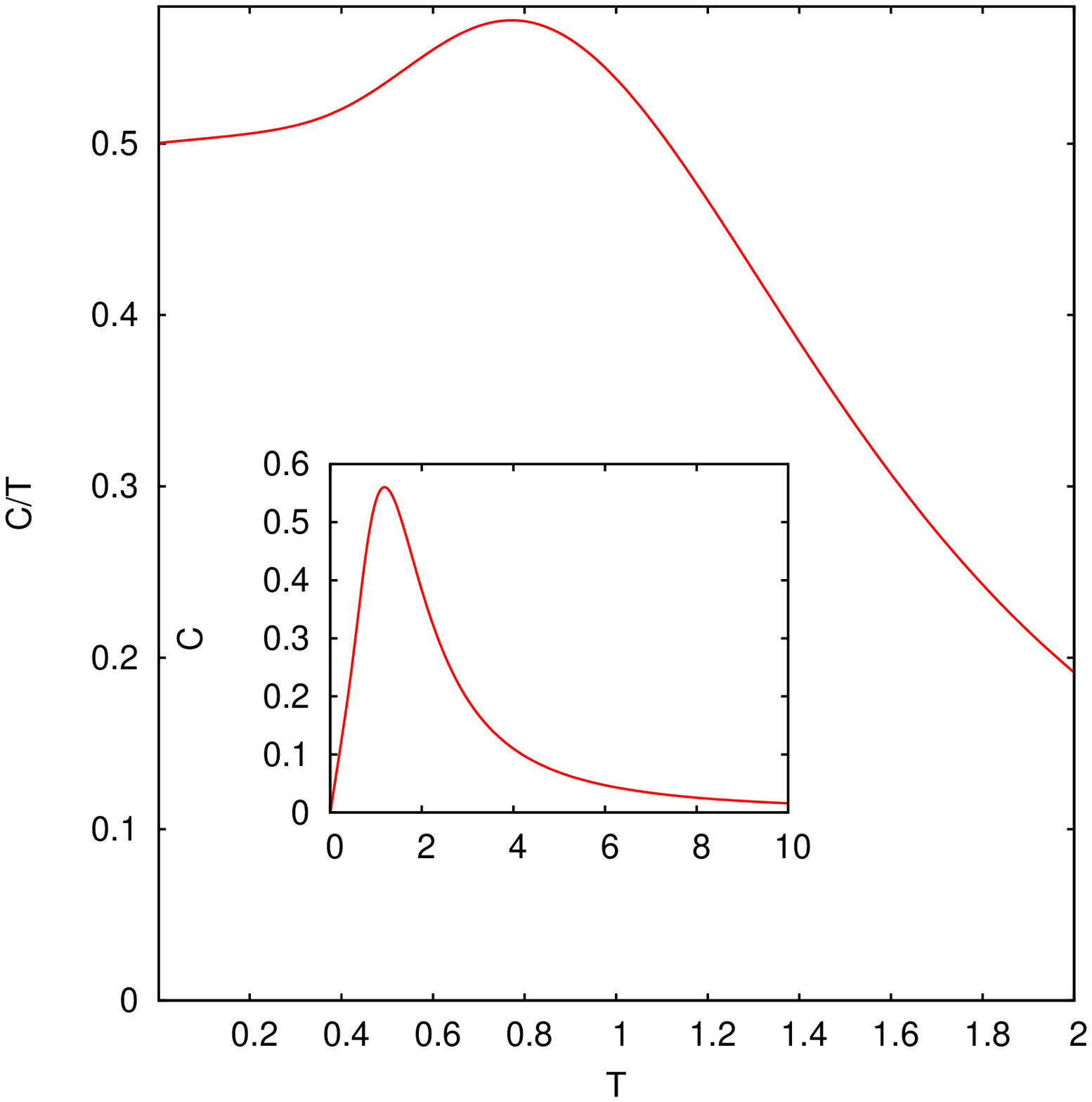}
\end{center}
\end{minipage}%
\begin{minipage}{0.5\linewidth}
\begin{center}
\includegraphics[width=\linewidth]{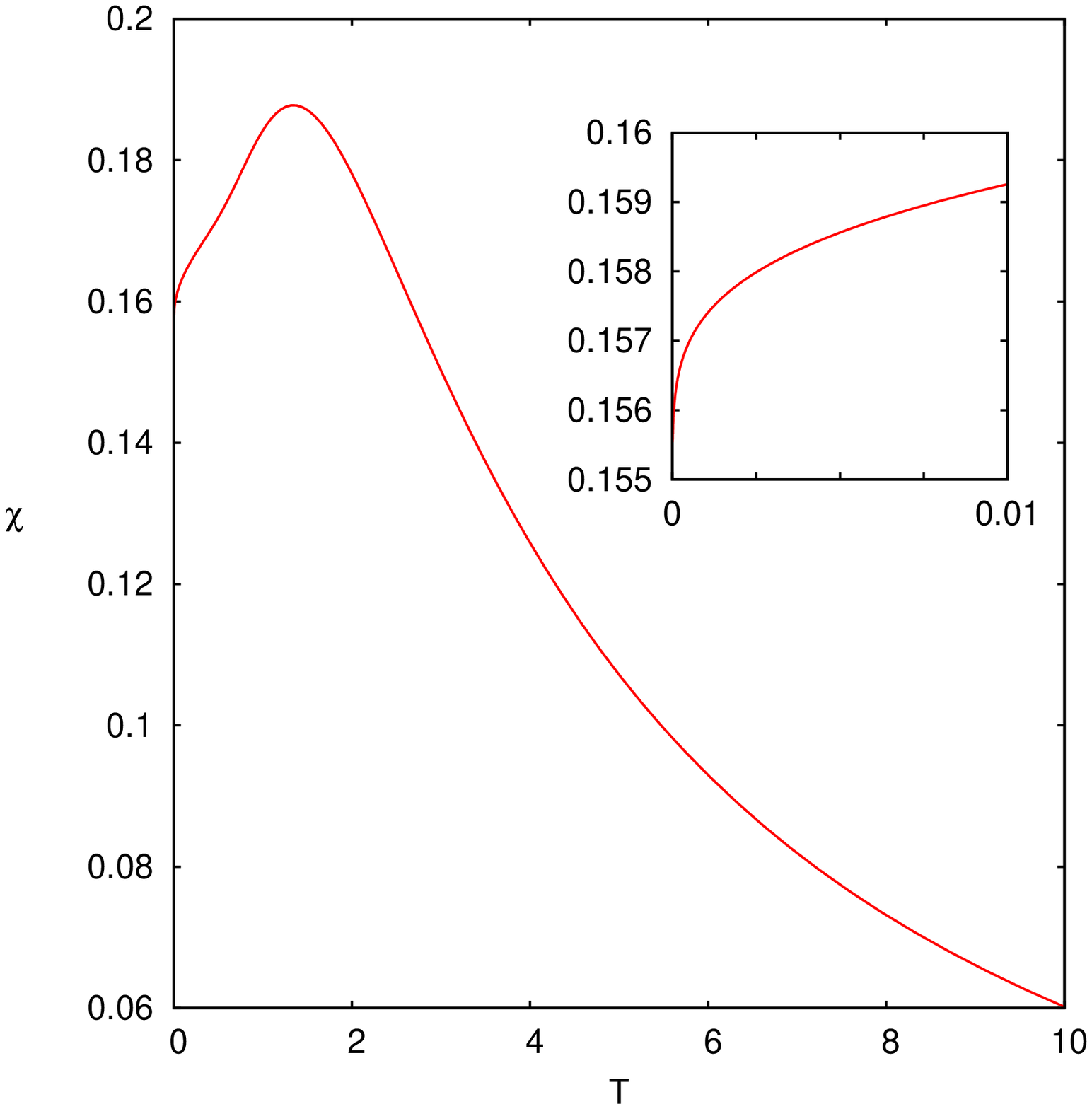}
\end{center}
\end{minipage}
\caption{On the left panel, we show the specific heat $C/T$ divided by temperature (in the inset we plot $C$ versus $T$) in units of $J=1$. On the right panel, we show the magnetic susceptibility as function of temperature and in the inset we show the existence of logarithmic corrections.}
\label{fig01}
\end{figure}

In order to illustrate, we can compute some thermodynamical quantities like the specific heat (at fixed chemical potential) and magnetic susceptibility (at zero field) out of the solution of the NLIE. The results are shown in the Figure \ref{fig01}. At very low temperatures the specific heat behaves linearly, revealing a gapless conformal spectrum. The slope $\frac{1}{2}$ is in agreement with our analytical calculation for the effective central charge given below. We also plot the magnetic susceptibility at zero magnetic field, right panel on Figure \ref{fig01}. The infinity slope at $T=0$ signals the existence of logarithm corrections.

\subsection{Low temperature behaviour and the effective central charge}

In order to study the low temperature behaviour, we may use the same trick as in the row-to-row transfer matrix case to evaluate the leading correction when temperature is finite but very close to zero. Once again, if we perform the change of variable $s=\pm (s'+\frac{3}{2 \pi} \ln \beta)$, we find the thermodynamical potential correction over the ground-state energy 
\begin{multline}
f(x)-e(x)=-\frac{1}{2 \pi \beta}\Bigg\{\int_{-\frac{3}{2 \pi} \log \beta}^{\infty} {\cal D}_+'(x-s'+\frac{\im}{4})  l\mathfrak{B}^+(s')
+ {\cal D}_-'(x+s'+\frac{\im}{4})  l\mathfrak{B}^-(s')
{\rm d}s'\\+
\int_{-\frac{3}{2 \pi} \log \beta}^{\infty} {\cal D}_+'(x-s'-\frac{\im}{4})  l\bar{\mathfrak{B}}^+(s')
+ {\cal D}_-'(x+s'-\frac{\im}{4})  l\bar{\mathfrak{B}}^-(s')
{\rm d}s'
\\+
\int_{-\frac{3}{2 \pi} \log \beta}^{\infty} {\cal D}_+'(x-s')  l Y_c^+(s')
+ {\cal D}_-'(x+s')  l Y_c^-(s')
{\rm d}s'\Bigg\},
\end{multline}
where
\begin{align}
l \mathfrak{b}^{\pm}(x)&:=\log \mathfrak{b}(\pm(x+\frac{3}{2 \pi}\log \beta))~~~~l \mathfrak{B}^{\pm}(x):=\log \mathfrak{B}(\pm(x+\frac{3}{2 \pi}\log \beta)), \nonumber\\
l \bar{\mathfrak{b}}^{\pm}(x)&:=\log \bar{\mathfrak{b}}(\pm(x+\frac{3}{2 \pi}\log \beta))~~~~l \bar{\mathfrak{B}}^{\pm}(x):=\log \bar{\mathfrak{B}}(\pm(x+\frac{3}{2 \pi}\log \beta)),\nonumber\\
l y^{\pm}(x)&:=\log y(\pm(x+\frac{3}{2 \pi}\log \beta))~~~~l Y^{\pm}(x):=\log Y(\pm(x+\frac{3}{2 \pi}\log \beta)), \nonumber\\
{\cal D}_{\pm}'(x)&:={\cal D}'(x\mp\frac{3}{2 \pi} \log \beta)\approx \frac{2 \pi}{\beta \sin (\frac{\pi}{3})}{\rm e}^{\pm\frac{2 \pi x}{3}}.\label{Aprox2}
\end{align}

Because of approximation (\ref{Aprox2}) we find
\begin{multline}
f(x)-e(x)=-\frac{1}{\beta^2 \sin(\frac{\pi}{3})}\Bigg\{ {\rm e}^{\frac{2 \pi x}{3} } \int_{-\frac{3}{2 \pi} \ln \beta}^{\infty}{\rm e}^{-\frac{2 \pi s}{3} } {\begin{bmatrix}{\rm e}^{\frac{\pi \im}{6}}\\{\rm e}^{\frac{-\pi \im}{6}}\\1 \end{bmatrix}}^t \cdot
\begin{bmatrix}l \mathfrak{B}^+(s) \\ l \bar{\mathfrak{B}}^+(s) \\ l Y^+(s) \end{bmatrix}
{\rm d}s
+\\
+
{\rm e}^{-\frac{2 \pi x}{3} } \int_{-\frac{3}{2 \pi} \ln \beta}^{\infty}{\rm e}^{-\frac{2 \pi s}{3} } {\begin{bmatrix}{\rm e}^{\frac{-\pi \im}{6}}\\{\rm e}^{\frac{\pi \im}{6}}\\1 \end{bmatrix}}^t \cdot
\begin{bmatrix}l \mathfrak{B}^-(s) \\ l \bar{\mathfrak{B}}^-(s) \\ l Y^-(s) \end{bmatrix}
{\rm d}s
\Bigg\},
\end{multline}
and the NLIE becomes 
\eq
\begin{bmatrix}
l\mathfrak{b}^{\pm}(s)\\
l \bar{\mathfrak{b}}^{\pm}(s)\\
l y_c^{\pm}(s)
\end{bmatrix}=
-\begin{bmatrix}
F_1 & F_2 & F_3\\
\bar{F}_2 & F_1 & \bar{F}_3\\
\bar{F}_3 & F_3 & F_4
\end{bmatrix} \ast 
\begin{bmatrix}
l \mathfrak{B}^{\pm}(s)\\
l \bar{\mathfrak{B}}^{\pm}(s)\\
l Y_c^{\pm}(s)
\end{bmatrix}
- \frac{2 \pi {\rm e}^{-\frac{2 \pi s}{3}}}{\sin(\frac{\pi}{3})}
\begin{bmatrix}
{\rm e}^{\pm \frac{\pi \im}{6}}\\
{\rm e}^{\mp \frac{\pi \im}{6}}\\
1
\end{bmatrix}. \label{NLIEscalingT}
\en
Now if we build the quantities
\eq
\Delta_{\pm}=\int_{-\frac{3 \ln \beta}{2 \pi} \approx -\infty}^{\infty}  {\begin{bmatrix}l \mathfrak{B}^{\pm}(s) \\ l \bar{\mathfrak{B}}^{\pm}(s) \\ l Y^{\pm}(s) \end{bmatrix}}^t \cdot {\begin{bmatrix}l \mathfrak{b}^{\pm'}(s) \\ l \bar{\mathfrak{b}}^{\pm'}(s) \\ l y^{\pm'}(s) \end{bmatrix}}
-{\begin{bmatrix}l \mathfrak{B}^{\pm'}(s) \\ l \bar{\mathfrak{B}}^{\pm'}(s) \\ l Y^{\pm'}(s) \end{bmatrix}}^t \cdot {\begin{bmatrix}l \mathfrak{b}^{\pm}(s) \\ l \bar{\mathfrak{b}}^{\pm}(s) \\ l y^{\pm}(s) \end{bmatrix}}
 {\rm d}s,
\en
we can show that the kernel contributions in (\ref{NLIEscalingT}) vanish away because of the symmetry $F_{ij}(x)=F_{ji}(-x)$, where $i$ and $j$ denotes the element position in Kernel matrix. We find
\begin{eqnarray}
f(x)-e(x)&\approx &-\frac{3}{8 \pi^2 \beta^2}\left({\rm e}^{\frac{2 \pi x}{3}} \Delta_+ + {\rm e}^{-\frac{2 \pi x}{3}} \Delta_-\right) \nonumber\\
&=& -\frac{3 \cosh(\frac{2 \pi x}{3})}{4 \pi^2 \beta^2}\left(L_+(\mathfrak{b}(\infty))+L_+(\bar{\mathfrak{b}}(\infty))+L_+(y_c(\infty))\right) \nonumber\\
&=&-\frac{3 \cosh(\frac{2 \pi x}{3})}{4 \pi^2 \beta^2}\left(2 L_+\left(\frac{1}{2}\right)+L_+\left(\frac{1}{3}\right)\right). 
\end{eqnarray}
Additionally to the functional relation (\ref{Functional}), Rogers dilogarithm function also satisfy\cite{JSUZUKI}
\eq
2 L_+\left(\frac{1}{n}\right)+\sum_{j=2}^n L_+ \left(\frac{1}{j^2-1}\right)=\frac{\pi^2}{6},
\en
therefore
\eq
f(x)-e(x)\approx-\frac{\cosh(\frac{2 \pi x}{3})}{4\beta^2} ~~ \Rightarrow~~ c_{eff}=\frac{3}{ 2 \pi} v_s=1,
\en
where the sound velocity is $v_s=\frac{2 \pi}{3}$ \cite{MARTINS1995}.

The fact that the effective central charge is positive guarantees that the specific heat and other thermodynamical quantities are also positive. By its turn, the negative central charge $c=-2$ obtained from the finite size analysis does not appear directly in the thermodynamical quantities, however the different central charges obtained from finite size and finite temperature analysis are known to be related as follows\cite{MARTINS1998}
\eq
c_{eff}=c+12 x_p,
\en
where in the case of $osp(1|2)$ model the lowest conformal dimension is $x_p=\frac{1}{4}$ \cite{MARTINS1998}.

\section{Conclusion}
\label{conclusion}

In this paper we derived non-linear integral equation either to the largest eigenvalue of the row-to-row transfer matrix or to the eigenvalue of the quantum transfer matrix. This allowed us to evaluate numerically the transfer matrix eigenvalue as a function of the system size and also the thermodynamical quantities like specific heat and magnetic susceptibility as a function of temperature. We have analytically obtained the (effective) central charge of the model from the derived non-linear integral equations, which is in agreement with the predicted results via numerical extrapolation of finite size data \cite{MARTINS1998}.

We expect that these results may be further extended to describe excited states and other quantum spin chain invariant by other superalgebras.

\section*{Acknowledgments}

G.A.P. Ribeiro thanks M.J. Martins, F. G\"ohmann, A. Kl\"umper for discussions. The authors are grateful for partial support by DFG through the program FOG
2316 and thank for the hospitality of Bergische Universit\"at Wuppertal where this work was completed.
T.S. Tavares thanks FAPESP for financial support through the grant 2013/17338-4. G.A.P. Ribeiro acknowledges financial support through the grants 2015/01643-8, S\~ao Paulo Research Foundation (FAPESP).

\end{document}